\documentclass[manuscript]{emulateapj}
\usepackage{times}
\usepackage{amsmath}
\usepackage[usenames,dvipsnames]{color}
\usepackage{multirow}

\newcommand{\myemail}{mathieu.powalka@astro.unistra.fr}

\slugcomment{To appear in Astrophysical Journal Letters}

\shorttitle{Globular cluster colors versus environment}
\shortauthors{Powalka et al.}

\begin{document}

\title{New constraints on a complex relation between globular cluster colors and environment} 

\author{Mathieu Powalka$^{1}$, Thomas H.~Puzia$^{2}$, Ariane~Lan\c{c}on$^{1}$, Eric W. Peng$^{4,5}$, Frederik Sch\"onebeck$^{3}$, 
Karla Alamo-Mart\'inez$^{2}$, Sim\'on \'Angel$^{2}$, John P. Blakeslee$^{6}$, Patrick C\^{o}t\'e$^{6}$, Jean-Charles Cuillandre$^{7}$, Pierre-Alain Duc$^{7}$, Patrick Durrell$^{8}$, Laura Ferrarese$^{6}$, Eva K. Grebel$^{3}$, Puragra Guhathakurta$^{9}$, S. D. J. Gwyn$^{6}$, Harald Kuntschner$^{10}$, Sungsoon Lim$^{4,5}$, Chengze Liu$^{11,12}$, Mariya Lyubenova$^{13}$, J. Christopher Mihos$^{14}$, Roberto~P.~Mu\~noz$^{2}$, Yasna Ordenes-Brice\~no$^{2}$, Joel Roediger$^{6}$, Rub\'en S\'anchez-Janssen$^{6}$, Chelsea Spengler$^{15}$, Elisa Toloba$^{9,16}$, Hongxin Zhang$^{2}$
}
\email{\myemail}
\altaffiltext{1}{Observatoire Astronomique de Strasbourg, Universit\'e de Strasbourg, CNRS, UMR 7550, 11 rue de l'Universit\'e, F-67000 Strasbourg, France}
\altaffiltext{2}{Institute of Astrophysics, Pontificia Universidad Cat\'olica de Chile, Av.~Vicu\~na Mackenna 4860, 7820436 Macul, Santiago, Chile}
\altaffiltext{3}{Astronomisches Rechen-Institut, Zentrum f\"ur Astronomie der Universit\"at Heidelberg, M\"onchhofstra\ss e 12-14, 69120 Heidelberg, Germany}
\altaffiltext{4}{Department of Astronomy, Peking University, Beijing 100871, China}
\altaffiltext{5}{Kavli Institute for Astronomy and Astrophysics, Peking University, Beijing 100871, China}
\altaffiltext{6}{Herzberg Institute of Astrophysics, National Research Council of Canada, Victoria, BC V9E 2E7, Canada}
\altaffiltext{7}{AIM Paris Saclay, CNRS/INSU, CEA/Irfu, Universit\'e Paris Diderot, Orme des Merisiers, F-91191 Gif-sur-Yvette Cedex, France}
\altaffiltext{8}{Department of Physics and Astronomy, Youngstown State University, One University Plaza, Youngstown, OH 44555, USA}
\altaffiltext{9}{UCO/Lick Observatory, Department of Astronomy and Astrophysics, University of California Santa Cruz, 1156 High Street, Santa Cruz, CA 95064, USA}
\altaffiltext{10}{European Southern Observatory, Karl-Schwarzschild-Str. 2, 85748 Garching, Germany}
\altaffiltext{11}{Center for Astronomy and Astrophysics, Department of Physics and Astronomy, Shanghai Jiao Tong University, Shanghai 200240, China}
\altaffiltext{12}{Shanghai Key Lab for Particle Physics and Cosmology, Shanghai Jiao Tong University, Shanghai 200240, China}
\altaffiltext{13}{Kapteyn Astronomical Institute, 9700 AV Groningen, The Netherlands}
\altaffiltext{14}{Department of Astronomy, Case Western Reserve University, Cleveland, OH, USA}
\altaffiltext{15}{Department of Physics \& Astronomy, University of Victoria, Victoria, BC, V8W 2Y2, Canada}
\altaffiltext{16}{Physics Department, Texas Tech University, Box 41051, Lubbock, TX 79409-1051, USA}

\begin{abstract}
We present an analysis of high-quality photometry for globular clusters (GCs) in the Virgo cluster core region, based on data from the Next Generation Virgo Cluster Survey (NGVS) pilot field,
and in the Milky Way (MW) based on VLT/X-Shooter spectrophotometry.~We find significant discrepancies in color-color diagrams between sub-samples from different environments,
confirming that the environment has a strong influence on the integrated colors of GCs. GC color distributions along a single color are not sufficient to capture the differences we observe in color-color space.
~While the average photometric colors become bluer with increasing radial distance to the cD galaxy M87, we also find a relation between the environment and the slope and intercept of the color-color relations.
~A denser environment seems to produce a larger dynamic range in certain color indices.~We argue that these results are not due solely to differential extinction,
IMF variations, calibration uncertainties, or overall age/metallicity variations.
~We therefore suggest that the relation between the environment and GC colors is, 
at least in part, due to chemical abundance variations, which affect stellar spectra and stellar evolution tracks.~Our results demonstrate that stellar population diagnostics derived from model predictions which are calibrated on one particular sample of GCs may not be appropriate for all extragalactic GCs.
~These results advocate a more complex model of the assembly history of GC systems in massive galaxies that goes beyond the simple bimodality found in previous decades.
\end{abstract}


\section{Introduction}

Globular clusters (GCs) come in different colors, which are low-resolution diagnostics of the rich collection of astrophysical parameters that characterize their constituent stellar populations.~It is well established that there are blue and red populations of GCs in every massive galaxy \citep[][]{zepf1993, peng2006}, which correspond to metal-poor and metal-rich stellar populations \citep[][]{puzia2005a, puzia2005b, colucci2009, colucci2014}, and that their proportion depends on the environment, in particular, the mass of the host galaxy \citep[][]{forbes1997, cote1998, gkp1999, larsen2001} and the galactocentric distance \citep[][]{geisler1996, harris2009, strader2011}.~In general, the GC system color distribution of a more massive galaxy will be broader and its mean shifted to redder colors than in a less massive one.~This trend is often interpreted as a radial metallicity gradient. It can be due to 1) the changing ratio of red and blue GCs and/or 2) the decreasing (i.e.~bluer) peak color of red {\it and} blue GC sub-populations as a function of galactocentric radius \citep[][]{harris2009, strader2011, oldham2016}.~In most previous studies, such analyses were based on a single photometric color of rich GC systems \citep{peng2006, peng2011, jordan2015}.~Those studies that used color-color planes as diagnostic tools were hampered by relatively small GC sample sizes to be able to assess any environmental dependence \citep[][]{puzia2002, hempel2004}.

In this letter, we present a detailed color-color plane analysis of the GC photometric properties in the pilot region of the {\it Next Generation Virgo Cluster Survey} \cite[NGVS/NGVS-IR, see][]{ferrarese2012, munoz2014}.~We compare various GC sub-samples located in different environments around the central massive cD Virgo galaxy M87 and in the MW.~While M87 constitutes one of the densest environments in the local universe \citep[$D\!\simeq\!16.5$\,Mpc, see][]{mei2007, blakeslee2009}, the MW environment shows a relatively shallow and smooth gravitational potential \citep{tully2015}.

\section[]{The data}
\label{sec_data}
\subsection[]{Next Generation Virgo Survey GCs}
\label{ngvsgcs}
The NGVS-GC sample used for our analysis is taken from \citep[][hereafter Paper I]{powalka2016}.
~It contains 1846 GCs within the 3.62~$\rm{deg}^2$ field around M87 (the Virgo core region) and provides photometric observations in $u^*$, $g$, $r$, $i$, $z$ and $K_s$ filters.
~This sample contains objects with {\sc SExtractor} magnitude errors smaller than 0.06 mag in each band, typical magnitudes around 21 in $i$, and typical masses of about $2\,\times\,10^6\,M_{\odot}$.
~Paper I provides limits on systematic photometric errors (of order 2-3\,\% in most bands, 5\,\% in $u$), and notes these would lead to global shifts in color-color diagrams.
Here, we apply the offsets $u_{\rm AB}\!=\!u_{\rm SDSS}-0.04$ mag and $z_{\rm AB}\!=\!z_{\rm SDSS}+0.02$ mag recommended in the Sloan Digital Sky Survey (SDSS) Data Release 10 \citep{ahn2014},
that were discussed but not applied in Paper I.~The GC sample was selected in a ``modified $uiK$ diagram",
that combines $(u-i)$ and $(i\!-\!K_s)$ color information with a compactness index measured on the NGVS $i$-band images (i.e.~structural information of the sources).
This ensures a very robust separation between GCs, stars and galaxies, in contrast to any separation one would obtain from optical colors alone.
The estimated contamination of the GC sample is of about 5\,\%, and is mainly due to stars at the blue end of the GC color distribution or to compact background galaxies.
~For all additional information, we refer the reader to Paper I.

\begin{figure*}[!t]
\begin{center}
\includegraphics[width=0.73\textwidth]{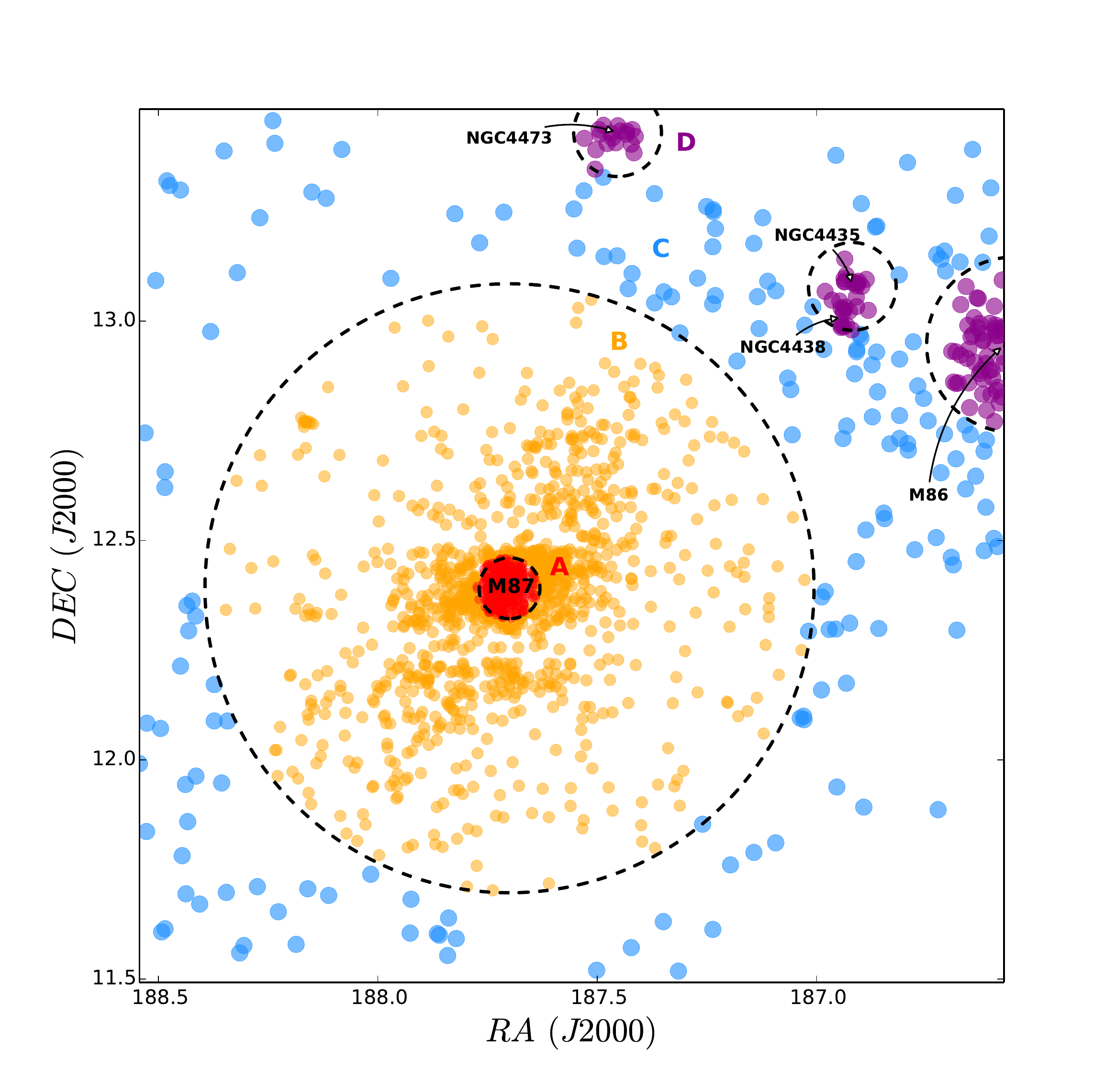}
\caption{\label{plot_sdss} Spatial distribution of the NGVS GC sub-samples. The red GCs are located within a projected radius of $20\!$ kpc from the M87 center.~The orange points show GCs between $20\!\leq\!r\!<\!200$\,kpc. The blue GCs are defined by $r\!>\!200\!$ kpc and being not associated with the M86, NGC\,4435, NGC\,4438 and NGC\,4473. GCs associated with the latter galaxies are indicated by magenta points.}
\end{center}
\end{figure*}

\begin{figure*}[!t]
\begin{center}
\includegraphics[width=19cm]{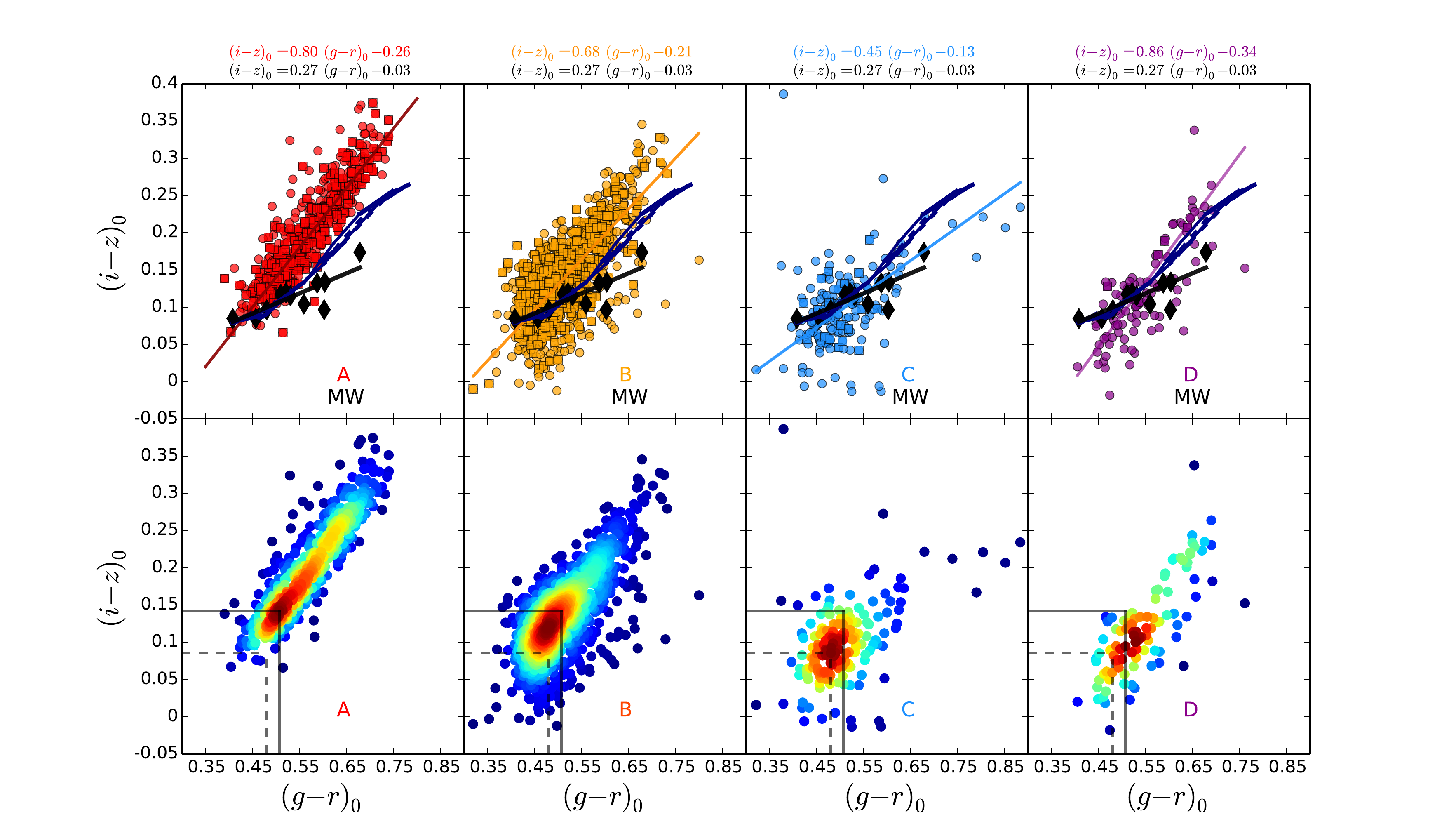}
\caption{\label{3ccd} $(g\!-\!r)_0$ versus $(i\!-\!z)_0$ color-color diagram for different GC samples.
({\it Top panels}): Comparison of various NGVS GC sub-samples (from left to right: {\bf A}, {\bf B}, {\bf C}, {\bf D}), with MW GCs (black diamonds).~Circles and squares mark GCs without and with radial velocity information.~Linear ML-fits to the NGVS-GC sub-samples and MW GCs are given in the top of each panel.~In addition, we show SSP predictions taken from the \citet{bc03} model for metallicities $0.0002<Z<0.03$ and ages $6<t<$13\,Gyr.~All colors are de-reddened with extinction values taken from \cite{schlegel1998}. ({\it Bottom panels}): Corresponding density plots for the NGVS GC sub-samples highlighting the colors of the highest-density peaks and their shift from the {\bf A} GCs toward the {\bf C} sample, illustrated by solid and dashed lines, respectively.}
\end{center}
\end{figure*}

\begin{figure*}[!t]
\begin{center}
\includegraphics[width=18.5cm]{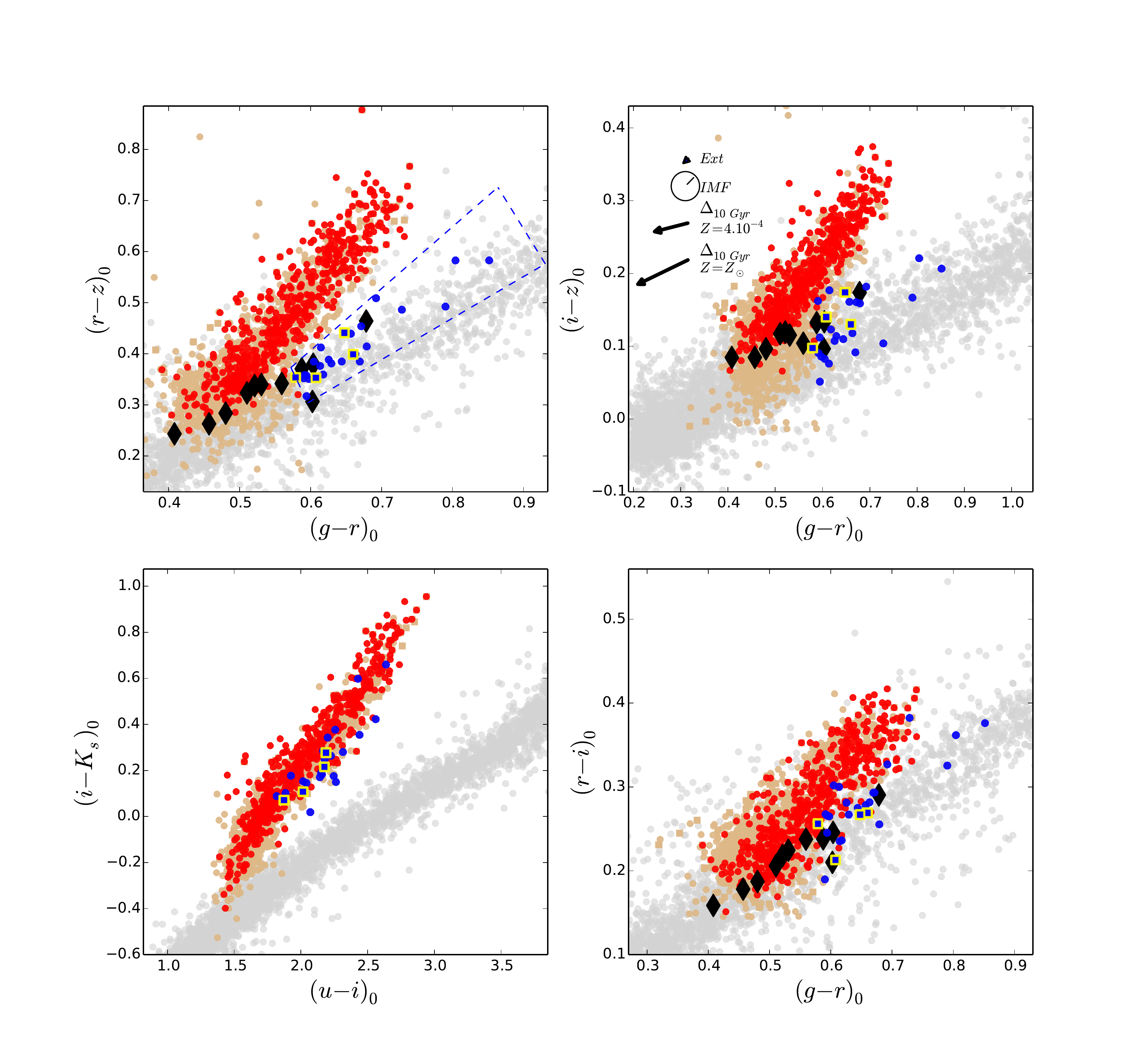}
\caption{\label{3ccdE} Color-color diagrams comparing the NGVS sample (colored dots, see Figure~\ref{plot_sdss} for the definition of subsets) and the MW GCs (black diamonds). GCs within $20$~kpc from the M87 center are highlighted in red and GCs from subset {\bf E} in blue, four of which are radial-velocity confirmed and marked with yellow contours.~The NGVS foreground stars are shown in grey.~Circles and squares mark GCs without and with radial velocity information consistent with Virgo cluster membership, respectively.~In the top right panel, vectors illustrate the shifts induced by several systematic changes.}
\end{center}
\end{figure*}

\begin{figure}[!t]
\begin{center}
\includegraphics[width=\columnwidth]{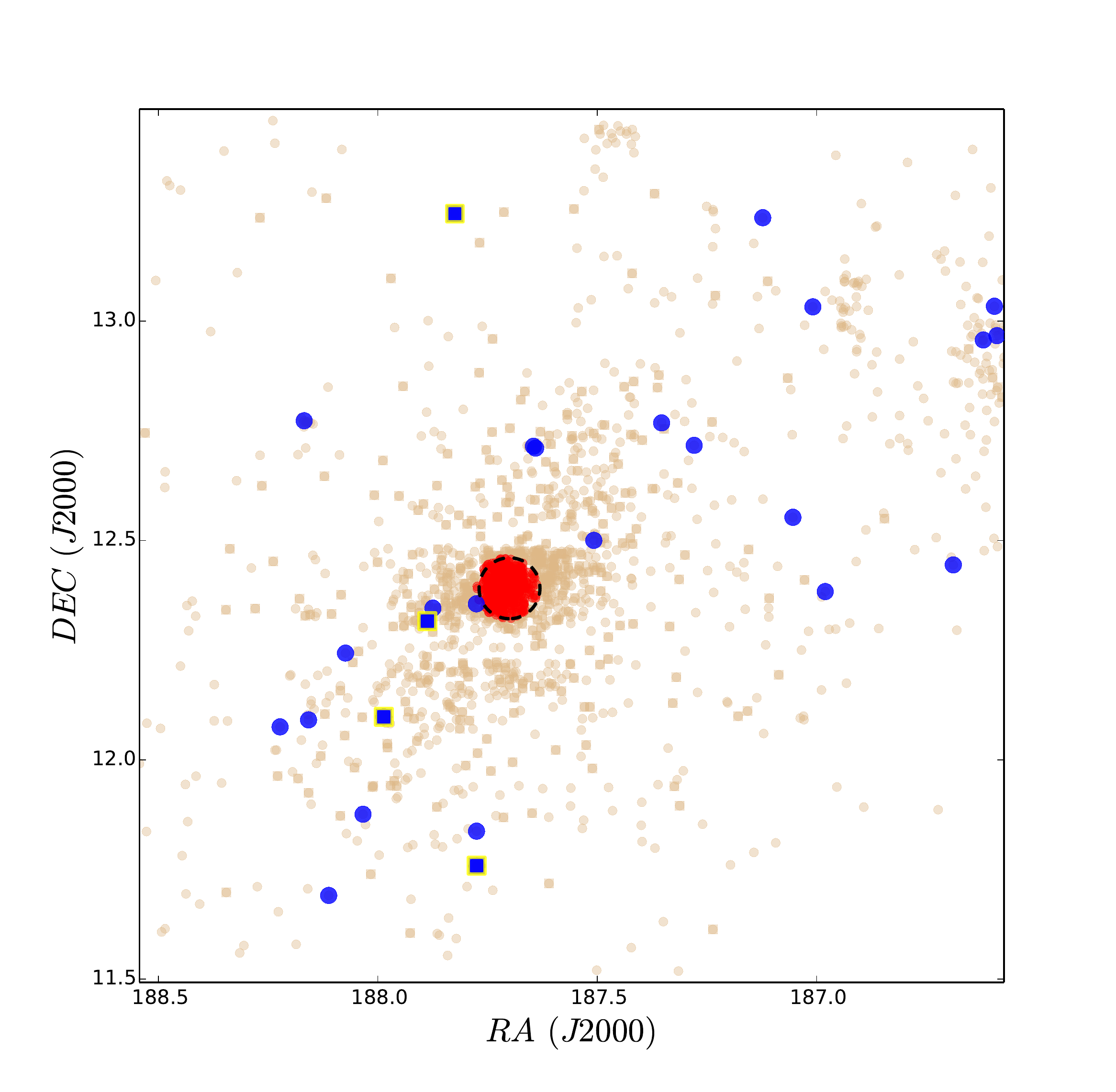}
\caption{\label{radeca1} Spatial distribution of the GC sub-samples defined in Fig\,\ref{3ccdE}.}
\end{center}
\end{figure}

\subsection[]{Milky Way GCs}
\label{mwgcs}

~Several photometric MW GC samples exist in the literature which are mainly based on optical Johnson-Cousins or SDSS photometry \citep[e.g.][]{harris2010, vdb14}.
Using such data in comparisons with MegaCam photometry requires transformation relations between systems.
Unfortunately, we found that the choice of a transformation relation 
and the internal uncertainties of these literature data introduce 
random and systematic uncertainties larger than the effects we 
wish to discuss, thus preventing a meaningful analysis.~A dedicated study, extended
to other galaxies, is postponed to a future article. 

~Pending further analysis of these photometric transformation uncertainties, 
 we favour the use of VLT/X-shooter spectra of MW GCs. 
Eleven are available to us, taken from the target sample of the 
{\it Panchromatic High-Resolution Spectroscopic Survey of Local Group Star Clusters} \citep[NGC\,104, NGC\,288, NGC\,362, NGC\,1851, NGC\,1904, NGC\,2298, NGC\,2808, NGC\,6656, NGC\,7078, NGC\,7089, and NGC\,7099; ][]{schonebeck2014}
~The spectra cover the near-UV to near-IR wavelengths and are calibrated to an absolute flux accuracy of better than $\sim$\,5\%, allowing for the computation of accurate synthetic colors
directly in the AB system of the NGVS data. 

X-shooter flux calibration errors occur on various scales (e.g.\ \citealt{moehler2014}).
~Errors on small scales 
are partly averaged out in broad band flux measurements,
resulting in magnitude errors below 0.01\,mag. Errors on larger spectral scales, or errors in the merging of data from two independent arms of the X-shooter instrument, can affect the colors more.
The 5\% bound applies to these (i.e.~+/-2.5\%).
A random distribution of possible large scale errors within the bounds leads to an estimated 1-sigma error of 0.02\,mag on colors (this has been tested by perturbing the GC calibration with
90 perturbation functions with a broad variety of shapes, all within the 5\% bounds).~We 
are aware of no reason that should induce color-dependent systematic errors on the color indices,
other than uncertainties in the transmission curves, and such errors are below 0.01\,mag (see Paper 1).

The metallicities of the Galactic GC sample 
span a range between $-2.3$ (NGC\,7078)$\,<\,$[Fe/H]$\,<\!-0.7$ (NGC\,104).
~The MW GC spectra have been obtained in drift-scan mode, i.e., the telescope was slewed across the clusters during the integration.
~Each GC was targeted with multiple scans at various locations, such that the total area covered by all scans corresponds to $\sim\!0.35\,\pi r_{\rm h}^2$ about the GC cluster center ($r_{\rm h}$ is the GC half-light radius).
~The sky subtraction was performed with dedicated sky drift-scans taken at positions typically $\sim\!1^\circ$ away from the GC centers.
For each cluster, all reduced scans have been stacked into a final spectrum that contains the luminosity weighted contributions of $\sim\!10^5$ GC stars.
A more detailed description of this data set will be presented in an upcoming paper (Sch\"onebeck et al.~2016, {\it in prep.}).

The synthetic colors of the MW clusters were computed with the transmission curves of \citet{betoule2013}, as recommended in Paper I.
Extinction corrections are based on the values of the McMaster catalog \citep{harris2010}.

\section{Results}
\label{results}
\subsection{The Influence of the Environment}
\label{inf_eff}
The NGVS pilot field includes several Virgo galaxies and their GC systems.~In Figure~\ref{plot_sdss}, we use the proximity to host galaxies to define four GC subsets.~Subset {\bf A} contains the GCs within 20\,kpc of M87, while subset {\bf B} covers the outer regions of M87 out to 200\,kpc ($r\!<\!41.6\arcmin$).~Sub-sample {\bf C} includes all the GCs located even further out from M\,87, with the exception of those located around other relatively large galaxies, which are grouped in subset {\bf D}.

We begin to analyze the optical $griz$ color-color distributions of these four subsets in Figure~\ref{3ccd}, and observe that they differ both in mean color and shape.~To guide the eye, and to recall the typical degeneracy between age and metallicity in the predicted colors of single stellar populations (SSP), we superimpose a set of models from \cite{bc03} with ages $6\!<\!t\!<\!13$\,Gyr and metallicities $0.0002\!<\!{\rm Z}\!<\!0.03$.~While this degeneracy is strong in all model sets with given abundance ratios, we caution that the actual loci and shapes of synthetic distributions remain strongly model-dependent, as illustrated extensively in Paper I.

Samples {\bf A}, {\bf B} and {\bf C} demonstrate that the $(g\!-\!r)_0-(i\!-\!z)_0$ color-color distribution changes significantly with distance to the center 
of M87.
This  is highlighted in the density plots of the bottom row of Figure \ref{3ccd}: for the peak of the distribution (usually referred to as the blue peak) we report color differences of $\Delta(i\!-\!z)\!\simeq\!0.06$ mag and $\Delta(g\!-\!r)\!\simeq\!0.03$ mag between subsets {\bf C} and {\bf A}.~Unlike sample {\bf A}, samples {\bf B} and {\bf C} exhibit a shallower color-color relation and are increasingly offset towards bluer average colors. This evolution is consistent with {\bf B} containing a composite of {\bf A} and {\bf C} GCs.~We note that the color-color relation of the MW GCs (black diamonds) best matches subset {\bf C}.

To characterize the influence of the environment, we have computed the maximum-likelihood linear relation between $(g\!-\!r)_0$ and $(i\!-\!z)_0$ for each GC sample.
These fitted lines mainly help emphasizing the overall trend of the color-color distribution.~Sub-samples {\bf D} and {\bf A} share similarly steep slopes, whereas the computed MW GC color-color slope is shallowest but similar to that of Virgo sample {\bf C}.~However, we caution that the MW GC sample contains only 11 data points.~Despite this 
limitation it can be stated that the GCs located nearest to M87 host stellar populations with significantly different properties than those in set {\bf C} or in the MW sample.

Differences between GC color distributions have been discussed mainly in terms of metallicity distributions in the past.~Our results indicate that extra parameters are required. Considering metallicity together with age remains insufficient, because of the strong degeneracy between age and metallicity predicted by old SSP models in the relevant color-color planes.~At least a third parameter is necessary.

\subsection{Comparison between Virgo and Milky Way GCs}
We present in Figure~\ref{3ccdE} three additional color-color diagrams ($grz$, $uiK$, and $gri$ along with the $griz$ plane from Figure~\ref{3ccd}) comparing NGVS GCs (colored dots) with the MW GCs (black diamonds).~The locus of M87 GCs (subset {\bf A}) is impressively tight, although the dispersion around this locus is slightly larger than what is expected from random photometric errors. In general, the color distributions of M87 and MW GCs are strikingly different, in particular in the $grz$ and $griz$ planes. The M87 clusters have redder $r\!-\!z$ and $i\!-\!z$ colors than MW clusters.~Moreover, the slopes of the trends differ for the two samples.

In the top right panel of Figure~\ref{3ccdE}, vectors show the shifts in the $griz$ plane resulting from: the average extinction vector; the change of the index of a power-law IMF from $-0.3$ to $-4.3$ (the vector depends on age and metallicity, hence the ellipse); an age difference of 10 Gyr at $Z\!=\!10^{-4}$, and the same at $Z\!=\!Z_\odot$.~None of these changes induces a variation that would reasonably explain the observations (see Section~\ref{discussion}).

Using the MW clusters in the $grz$ diagram to guide the eye, we find that the Virgo sample contains a subset of some 30 red GCs whose colors align with those of the MW GCs, rather than with those of M87. We select the most obvious of these in the $grz$ panel of Figure~\ref{3ccdE} (inside the blue polygon) and display them as blue dots in the other panels, referring to these as subset {\bf E}\footnote{We have removed five objects from the initial subset {\bf E}, of which we suspect three may be affected by dust lanes, and two might be background galaxies due to their elongated shape.~A careful visual inspection confirms that all the remaining candidates (26) have apparently normal GC properties.}.~It is worth noting that these particular clusters, like any other cluster with colors similar to those of MW GCs, would have been easily mistaken for stars on the basis of optical colors alone (grey dots in Figure~\ref{3ccdE}).~The $uiK$ diagram, however, clearly separates them from stars, as do their slightly extended radial profiles, which are akin to typical GCs, but about $2\times$ narrower than the ones expected for Ultra Compact Dwarfs (UCDs) at the Virgo distance.~Moreover, four of the subset {\bf E} objects are spectroscopically confirmed as GCs, whereas no data is available for the other 22 candidates.

The subset {\bf E} GCs are uniformly spread over the NGVS pilot field, with some of them clustered around M86 and NGC\,4438 (Figure \ref{radeca1}).~We do not observe any overdensity around M87, which strongly suggests that they were formed in an environment other than the M87 host halo.~In a recent study, \citet{ferrarese2016} estimated that a significant fraction of the GCs in the Virgo core may be inherited from infalling galaxies, that were themselves fully or partly shredded by tidal forces.~It is thus plausible that the GCs identified here are born in an environment less dense than the Virgo core.

\section{Discussion}
\label{discussion}

Our data for Virgo and MW GCs suggests that the dependence of GC color distributions on environment is more complex than previously thought. In particular, their discussion cannot be restricted to the existence or absence of a bimodal metallicity distribution.~GCs in different environments may have color-color distributions that differ in slope in addition to being offset from each other. This result puts the definition of the usual blue and red sub-populations in question, and calls for a larger variety of GC formation scenarios.

Before examining physical parameters that may play a role in this complexity, we examine and eliminate potential observational biases.

\noindent {\it Photometry}: Systematic errors in the photometry would not explain differences in the shapes of the distributions, nor would they produce differences between the subsets we have defined within the NGVS sample.~A size-color relation is known among GCs in various galaxies based on Hubble Space Telescope data \citep[e.g.][]{jordan2005, puzia2014}, with blue clusters typically being $\sim\!20\%$ larger than red ones.~If the aperture corrections applied to NGVS clusters left size-dependent effects in the colors, we would expect those to be strongest at the blue end of the distribution, while we observe the largest internal deviations at the red end.~There is also a known size-luminosity relation for bright GCs in massive galaxies \citep[][]{puzia2014} and we might expect the color-color relations to deviate between bright and faint GCs.~No such trend was found in our data either.~Finally, we have checked that the trends we describe within NGVS are not residual effects of seeing differences between the individual fields of view combined to cover the Virgo core region. 

\noindent {\it Dust Extinction}: As the Virgo core region is located at high galactic latitude ($b\!=\!74^{\rm o}$), the extinction corrections are small \citep{schlegel1998}:~$\langle E_{\rm (B-V)}\rangle\!=\!0.0246$ mag and $\sigma_{\rm E_{\rm (B-V)}}\!=\!0.0037$ mag across the field. A change of the extinction law or a rescaling of the extinction vector would shift the color distributions without much changing their shape. We have found no spatial correlation between color-based subsets of clusters and the extinction map of \citet{schlegel1998}.

\noindent {\it Size effects in the MW GCs}: The observation areas of the MW GCs are restricted to $\sim\!0.35\,\pi r_{\rm h}^2$ around their respective centers.~Efficient mass segregation could have raised the relative number of massive stars around the center, compared to the cluster as a whole.~A changed proportion of a certain type of stars (e.g. main sequence, blue straggler or red giant branch) may cause a non-negligible color variation.~We tested this by adding stellar spectra of relevant types to the X-shooter spectra of the clusters.~The direction found for these variations is always roughly parallel to the MW sequence in the $griz$ plane and, therefore, cannot explain the difference with the GC locus around M87.
\smallskip

Having found no observational bias to explain the observed color distributions, we consider physical causes. As age and metallicity are highly degenerate in the $griz$ diagram, we consider parameters other than these two.

\noindent {\it GC mass}: MW GCs are typically about $10\times$ less massive than those in the Virgo sample. If GC mass was driving the color differences in the $griz$ diagram, the Virgo clusters in subset {\bf E} would be expected to have systematically lower masses than the other Virgo GCs in our dataset.

\begin{figure}[!b]
\begin{center}
\includegraphics[width=\columnwidth]{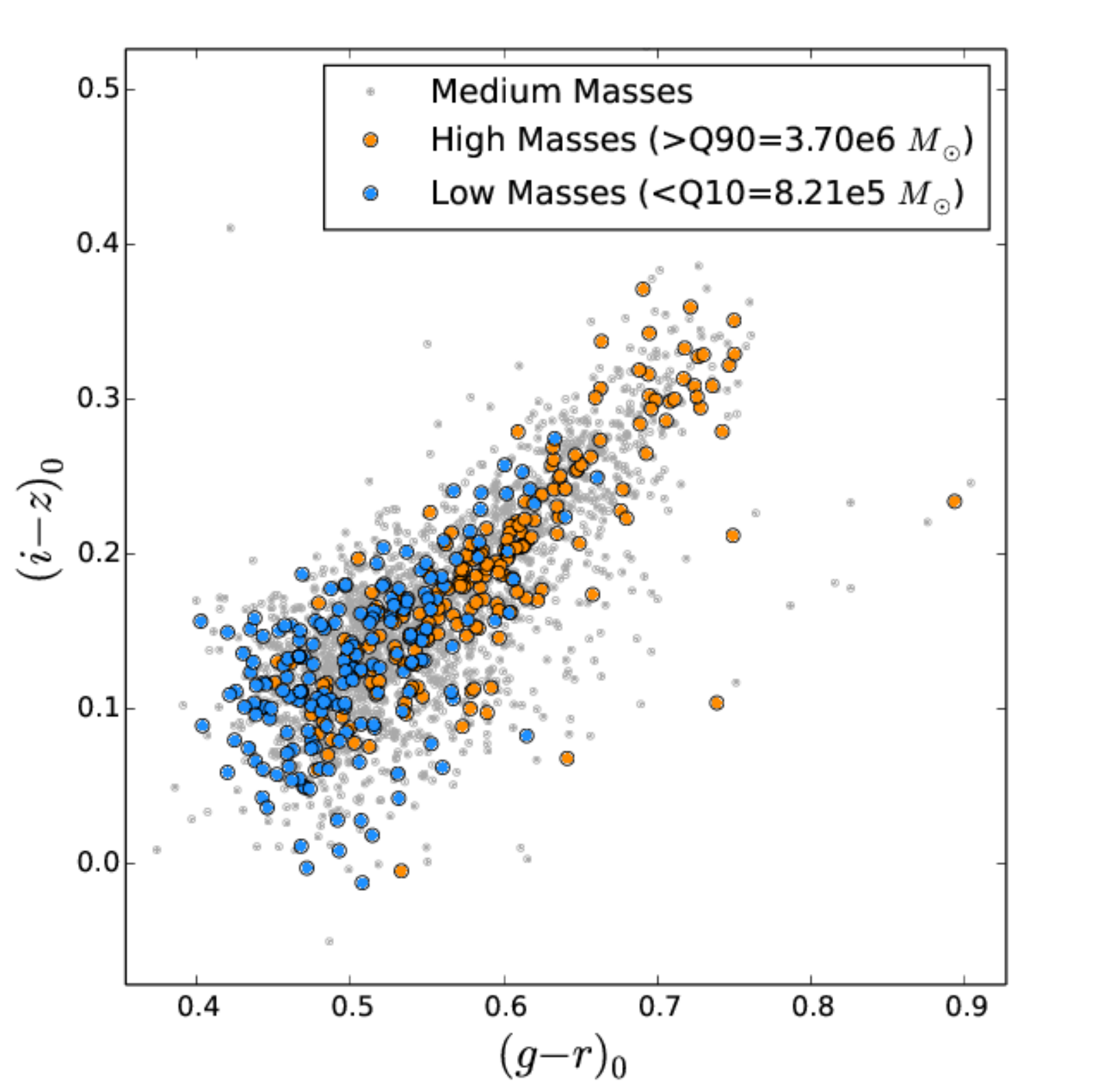}
\caption{\label{fig5} $(g\!-\!r)_0$ versus $(i\!-\!z)_0$ color-color diagram for three GC samples selected by their masses which provides evidence that mass is not the driving factor in the difference between Virgo GCs in subset {\bf E} or elsewhere.}
\end{center}
\end{figure}

We have estimated Virgo GC masses using SSP model inversion using predictions of 7 recent models (Powalka et al., {\it in prep}).~The mean estimated masses of each of our Virgo subsets are yet similar ($2.2\,\times\,10^6\ M_{\odot}$ for {\bf A}, {\bf D} and for the clusters isolated as subset {\bf E} in $griz$ and $1.8\,\times\,10^6\ M_{\odot}$ for {\bf B} and {\bf C}).~In addition, we show in Figure~\ref{fig5} two extreme 10\%-iles of the Virgo GCs mass distribution, with masses $\leq\!8.2\,\times\,10^5\ M_{\odot}$ (Q10, similar to the MW GCs) and $\geq\!3.7\times10^6\ M_{\odot}$ (Q90).~Although there is a lack of low-mass Virgo GCs at red colors\footnote{This is due to the too faint $u$-band fluxes of such low-mass, metal-rich GCs, which do not pass our photometric quality selection criteria.},~we find no significant correlation between GC mass (or luminosity) and the association with one or the other color-color sequence.~This implies that the GC colors are mainly influenced by the global environment, rather than by the local environment set by the GC mass.~However, a complete low-mass GC sample would be necessary to perfectly clinch this point.

\noindent {\it Initial Mass Function (IMF)}: Models constructed with a range of extreme IMFs produce a marginal modification (see Figure~\ref{3ccdE}), dwarfed by the age and/or metallicity variations expected for the GC samples \citep[][]{puzia2006}.

\noindent {\it Chemical Variance}: At this stage of the analysis, we attribute 
much of the dependence between the GC color-color locus and environment to abundance variations.~In massive MW clusters, detailed studies have shown that stars with a large range of chemical abundances may coexist, even when heavy-element stellar abundances are homogeneous (see e.g.~the recent review of \citealt{renzini2015}). Several previous studies have also hinted at abundance variations in M87 GCs \citep[][]{sohn2006, kaviraj2007, bellini2015}.~Abundance ratios influence the morphology of the horizontal branch (HB), which strongly influences the relation between optical and near-infrared colors \citep{conroy2010, maraston2011}.~However, the tight locus of our Virgo GC sample in the $uiK$ diagram is incompatible with wild variations of the HB morphology between clusters (unless some other parameters conspire to counteract the effect of the HB variations).~A similar argument also tends to exclude largely varying proportions of blue stragglers.

To explain our observations, a spectral effect localized in the range of the $r, i, z$ bands would be more suitable. Molecular bands that depend on the surface abundances of CNO-cycle elements, may produce such an effect.~Unfortunately, very few SSP models allow for CNO abundance variations, mostly because stellar spectral libraries are incomplete.~The computations recently started by \cite{aringer16} are, as yet, sampled too sparsely (e.g in metallicity and gravity for N-enhanced models) to conclude whether or not CNO abundances produce the required changes in color.~Models that consistently vary light element abundances both in the stellar evolution tracks (HB morphology) and in the stellar spectra (molecular bands) are lacking.~Finally, varying [$\alpha$/Fe] ratios might also play a role, for instance via molecular bands and the near-infrared CaII triplet.~Self-enrichment via core collapse supernovae is very limited in MW clusters \citep{renzini2015}, but the story may be different at the masses of the Virgo GCs studied here, as suspected in several studies \cite[e.g][]{mieske2006}.~We have briefly assessed this point using SSP model predictions from PEGASE \citep[][modified by MP, AL, and P. Prugniel]{leborgne2004} with two stellar libraries at [$\alpha$/Fe]~$\!=\!0.0$ and 0.4\,dex. The increase of [$\alpha$/Fe] produces a shift in the model predictions qualitatively from the MW towards the M87 GC sequence, but the amplitude of this shift is $4\times$ smaller than the observed offset. Other studies such as \citet{lee2009} found similar color offset amplitudes.

\section{Conclusion}
\label{Conclusion}

In this letter, we conduct a comparison of the optical color-color properties of GC samples from different environments.
~We find that previous descriptions of the effects of environment on GC color-distributions are insufficient to capture the actual diversity seen in color-color planes:
the samples studied exhibit separate color-color relations unexplained by the commonly accepted age and metallicity variations.

With the environmental subdivision in the NGVS pilot field, we observe that the Virgo subset {\bf A} (i.e.~GCs within 20\,kpc of M87) exhibit a steeper color-color relation than
the MW and the Virgo {\bf C} sample (i.e.~GCs located far from massive Virgo galaxies).

We note that a reduced subset of the NGVS GCs shares a color trend with the MW GCs. The spatial distribution of this subset lends credence to differences related to the environment.
~However, a spectroscopic confirmation of a larger sample would be strongly desirable to support the identification of this subpopulation.

We also confirm a relation between the mean GC color and the galactocentric distance to M87.
~We find a shift towards redder average colors with decreasing galactocentric radius, $\simeq\!2$ times larger in $(i\!-\!z)_0$ than in $(g\!-\!r)_0$,
although they share similar dynamic color ranges (i.e.~$\Delta{\rm color}/\sigma_{\rm color}$).

Finally, we show that photometric calibration, dust extinction, GC mass or IMF variations are unable to explain the observations.

A possible explanation for the measured color-environment correlations might be the imprint of global elemental abundance variations in the stellar atmospheres of GC stars.
~The lack of correlation with GC luminosity and mass of the variance in the color-color relations implies that such changes might not be due to the local GC environment,
but should have their cause in the global host galaxy environment.

To conclude, we believe that the complex relation between environment and chemical enrichment of GC populations could be a major constraint on galaxy formation models in the future decades.~This result could modify our current vision of the formation and assembly of GC systems, mainly by considering GCs conjointly with their host galaxy.~A more precise and quantitative assessment will be needed in the future
to deepen our understanding of these observations. 

\acknowledgments
This project is supported by FONDECYT Regular Project No.~1161817 and BASAL Center for Astrophysics and Associated Technologies (PFB-06).~We thank P. Prugniel for his help with the non-standard PEGASE calculations.

{\it Facilities:} \facility{CFHT (MegaCam/WIRCAM)}, \facility{VLT:Kueyen (X-shooter)}.

\clearpage

\end{document}